\documentclass[aps,prd,twocolumn,showpacs,floatfix,superscriptaddress,
preprintnumbers,nofootinbib]{revtex4}

\usepackage{graphicx,amssymb,natbib}
\usepackage[dvips]{color}
\definecolor{Black}{named}{Black}
\definecolor{Red}{named}{Red}
\definecolor{Blue}{named}{Blue}

\newcommand{\be}{\begin{equation}}
\newcommand{\ee}{\end{equation}}
\newcommand{\ba}{\begin{eqnarray}}
\newcommand{\ea}{\end{eqnarray}}

\def\lsim{\raise0.3ex\hbox{$\;<$\kern-0.75em\raise-1.1ex\hbox{$\sim\;$}}}
\def\gsim{\raise0.3ex\hbox{$\;>$\kern-0.75em\raise-1.1ex\hbox{$\sim\;$}}}

\def\theta{\vartheta}
\def\ap{\approx}

\def\rp{{\cal F}_{\bar p}/{\cal F}_{\bar p+p}}
\def\re{{\cal F}_{e^+}/{\cal F}_{e^++e^-}}

\begin{document}

\title{Antimatter production in supernova remnants}

\author{M.~Kachelrie\ss}
\affiliation{Institutt for fysikk, NTNU, Trondheim, Norway}

\author{S.~Ostapchenko}
\affiliation{Institutt for fysikk, NTNU, Trondheim, Norway}
\affiliation{D. V. Skobeltsyn Institute of Nuclear Physics, Moscow State University, Russia}

\author{R.~Tom\`as}
\affiliation{II. Institut f\"ur Theoretische Physik,
    Universit\"at Hamburg, Germany}

\begin{abstract}
We calculate the energy spectra of cosmic rays (CR) and
  their secondaries produced in a supernova remnant (SNR), taking into
  account the time-dependence of the SNR shock. We model the
  trajectories of charged particles as a random walk with a prescribed
  diffusion coefficient, accelerating the particles at each shock
  crossing. Secondary production by CRs colliding with gas is included
  as a Monte Carlo process. We find that SNRs produce less antimatter
  than suggested previously: The positron/electron ratio $\re$ and the
  antiproton/proton ratio $\rp$ are a few percent and ${\rm few}\times
  10^{-5}$, respectively. Moreover, the obtained positron/electron
  ratio decreases with energy, while the antiproton/proton ratio rises
  at most by a factor of two above 10\,GeV. 
\end{abstract}

%\keywords{cosmic rays: antimatter -- acceleration of particles -- 
%supernova remnants}
\pacs{98.70.Sa, %Cosmic rays 
      95.30.Cq  %Elementary particle processes
}

\maketitle

%%%%%%%%%%%%%%%%%%%%%%%%%%%%%%%%%%%%%%%%%%%%%
\section{Introduction}
%%%%%%%%%%%%%%%%%%%%%%%%%%%%%%%%%%%%%%%%%%%%%
Measurements of the antimatter fraction of cosmic rays (CR) provide
not only insight into CR physics itself \citep{galprop}, as e.g.\ 
their propagation in the galaxy, but are also valuable
probes for cosmology and particle physics. In particular, the
annihilation of dark matter (DM) leads to an equal injection
rate of matter and antimatter particles into the Galaxy, while the CR flux 
from astrophysical sources is matter-dominated. A possible
way to detect DM is therefore to estimate carefully the expected
antimatter fluxes from astrophysical sources and to search then for any
excess \citep{DM}.

The PAMELA collaboration presented recently results of their measurement 
of the positron fraction in CRs, which is rising rapidly from
10 to 100\,GeV \citep{PAMELA}. 
At the same time, the antiproton ratio measured by PAMELA declines
above 10\,GeV \citep{Adriani:2008zq}, consistent with expectations.
The conventional estimate for antimatter fluxes from 
astrophysical sources uses as only production mechanism of antimatter
the scattering of CRs on interstellar gas \citep{galprop}. As discussed
e.g.\  in \cite{Serpico:2008te}, the energy dependence of the
Galactic diffusion coefficient, $D\propto E^\delta$ with 
$\delta=0.5-0.6$, is inconsistent with an increase of the antimatter fraction
with energy. By contrast, the spectral shape of fragmentation functions 
leads quite naturally to such a rise in the case of DM annihilations
or decays.

The DM interpretation of the PAMELA excess faces however several 
difficulties \cite{DM}: First, the required rate of positron production is 
larger than expected for a stable thermal relic. As a consequence, either
the annihilation rate has to be enhanced by the clumpiness of
DM or by non-perturbative effects operating at small velocities, or
the DM particle should be unstable with the appropriate life-time.
Second, in gauge boson or quark fragmentation, positron, antiproton and
photon production are tied together and thus one has to postulate a DM particle
annihilating only into electrons and muons. More importantly, assuming
antimatter production by diffusing CRs as the only astrophysical source for
antimatter falls short: Since electrons lose fast energy, the high-energy
part of the $e^-+e^+$ spectrum should be dominated by local sources as 
nearby pulsars, as pointed out already 20~years ago~\citep{pulsar}. 
Moreover, 
electromagnetic pair cascades in pulsars 
result naturally in a large positron fraction together with a ``standard'' 
antiproton flux.

More recently, supernova remnants (SNR) were put forward as an alternative 
astrophysical explanation for a rising positron fraction \citep{B09}: 
Positrons created as secondaries\footnote{Note that in the literature
one often uses the notion of primary CRs in a more general sense for
all particles produced by CR sources.  Here we refer to the products 
of proton interactions in the source as secondaries.} 
of hadronic interactions in the shock 
vicinity participate in the acceleration process and, according to 
\cite{B09},  should thus have a flatter energy spectrum than primary 
electrons.  It was estimated 
that the resulting positron fraction can explain the PAMELA excess and 
rise up to 50\% at higher energies \cite{B09}, while subsequently a 
similar mechanism for antiprotons was 
suggested in~\cite{pp}. Since %diffusive 
shock acceleration in SNR is 
expected to be the main source for  Galactic CRs~\cite{SNR}, such
a scenario has also important consequences for the interpretations of CR data
as, e.g., the boron-to-carbon ratio \citep{BC}.
Additionally, several alternative explanations for the positron excess 
have been suggested: An inhomogeneous distribution of CR sources in the
solar neighbourhood was put forward in Ref.~\cite{piran},
while the authors of Ref.~\cite{coocon} proposed an enhanced secondary 
production in a cocoon-like region surrounding CR sources.

The present work examines the production of secondary $\bar p$ and $e^+$
in SNRs, improving on previous studies~\cite{B09,pp,sarkar} in two respects:
First, we use a Monte Carlo (MC)
approach calculating the trajectory of each particle individually in
a random walk picture. This makes it easy to include interactions and
the production of secondaries. Second, our approach allows 
us to include the time (and spatial) dependence of relevant parameters
describing the evolution of a SNR as, e.g., the shock radius and its 
velocity, the magnetic field
or the CR injection rate  and to test their influence on the CR spectra.
We should also stress what are {\em not\/} the aims of the present 
work: We do neither  address the problem of acceleration from a
microscopic point of view nor consider any feedback of CRs on
the shock or the magnetic field. Although the latter processes are 
important to obtain accurate CR escape fluxes, we shall show that our
simplified treatment leads to an upper limit on the secondary fluxes.
Moreover we restrict ourselves to the production of antiprotons and 
positrons in the source, being the key issue in the proposal
of Ref.~\cite{B09}, while we do not discuss secondary production 
during CR propagation.

%%%%%%%%%%%%%%%%%%%%%%%%%%%%%%%%%%%%%%%%%%%%%
\section{Simulation procedure}
%%%%%%%%%%%%%%%%%%%%%%%%%%%%%%%%%%%%%%%%%%%%%
Shocks around SNRs are supposed to be collisionless, with charged particles
scattering mainly on inhomogeneities of the turbulent magnetic field. We model 
such trajectories by a random walk in three dimensions with step size $l_0(E)$ 
determined by an energy-dependent diffusion coefficient $D$. Diffusion close 
to the shock is usually assumed to proceed in the Bohm regime with the
mean free path $l_0$ proportional to the Larmor radius $R_L$. Thus 
$D(E)\propto E$ and 
\be
 D = \frac{cl_0}{3} = f_B^{-1}\;\frac{c^2 p}{3eB} \,,
\ee
where $f_B$ denotes the ratio of the energy density in the turbulent and 
in the total magnetic field.
We neglect the coupling between CRs and the turbulent magnetic field,
assuming that a layer with Bohm diffusion extends far enough into the 
up-stream region. For a constant magnetic field, most low-energy
CRs do not escape but are confined in the SNR,
corresponding to an ``age-limited'' scenario for the CR flux from SNRs.

We describe the evolution of the shock in the rest frame of the SNR. 
Then the (yet unshocked) up-stream region is at rest, $v_1=0$, and has 
the density of the surrounding interstellar medium (ISM), 
$\rho_1=\rho_{\rm ISM}$. Assuming a strong shock 
with Mach number ${\cal M}\gg 1$, the shocked down-stream region 
flows with the velocity $v_2=3v_{\rm sh}/R$ and has the density $\rho_2=R\rho_1$.
Here, $R$ denotes the compression ratio $R=(\gamma+1)/(\gamma-1)=4$ 
for a mono-atomic gas with $\gamma=5/3$. We account for the flow,  
adding in the down-stream
region on top of the random walk an ordered movement of the particle
with velocity $v_2$ that is directed radially outwards. Thus a particle
trajectory evolves during the time step $\Delta t=l_0/c$ as
\be
 {\bf x}(t+\Delta t) = 
 {\bf x}(t)+ v_2\, \Delta t \,\theta(r_{\rm sh}-r)\, {\bf e}_r +{\bf l}_0 \,,
\ee
where ${\bf l}_0$ denotes a random step, $r_{\rm sh}$ the radius of the
spherical shock front at time $t$, and $\theta(x)$ the step function.

Crossing the shock, particles are accelerated. We neglect that the 
relative energy gain $\xi=(E_{k+1}-E_k)/E_k$ per cycle $k$ depends on the angle 
of the  trajectory to the shock front, and use for simplicity that on average 
for a non-relativistic shock
%
%\be
 $\xi=\frac{4}{3c}(v_2-v_1)=v_{\rm sh}/c$.
%\ee
%
For the position $r_{\rm sh}$ and the velocity $v_{\rm sh}$ of the
SNR shock we use the  $n=0$ case of the analytical solutions 
derived by \cite{TM99}. These solutions connect smoothly
the ejecta-dominated phase with free expansion $r_{\rm sh}\propto t$
and the Sedov-Taylor stage $r_{\rm sh}\propto t^{2/5}$. The acceleration of 
CRs is assumed to cease after the transition to the radiative
phase at the time $t_{\max}$.

As the injected particles diffuse, electrons lose energy via synchrotron 
radiation, while protons 
can scatter on gas of the ISM producing secondaries
that include antiprotons and positrons. Cross sections and the final state
of pp-interactions are simulated using QGSJET-II \cite{qgsjet}, 
while we use SIBYLL~2.1 \cite{sibyll1.7} for decays of unstable particles.

The last ingredient for our simulation procedure is an injection model. 
To ease the comparison with the  results of \cite{B09},
we fix the electron/proton ratio $K_{ep}$ at injection to 
$K_{ep}=7\times 10^{-3}$. As injection energy we use $E_0=10$\,GeV. 
In the first model used, the injection rate 
\be 
 \dot N \propto 
 r_{\rm sh}^2 \,v_{\rm sh}^\alpha \,\delta(E-E_0)\,\delta(r-r_{\rm sh}) 
\ee
is proportional to the volume swept out per time by the shock, i.e.\ 
$\alpha=1$ (thermal leakage model of Ref.~\cite{in1}).
In the second model, the injection rate $\dot N$ is 
proportional to the CR pressure \citep{in2} and $\alpha=3$.
In the case of model 2, the fraction of particles injected very early
is significantly larger than in model 1.

We use the following parameters to describe the SNR: We choose the
injected mass as $M_{\rm ej} = 4M_\odot$, the mechanical explosion energy 
as $E_{\rm snr} = 5\times 10^{51}$\,erg, and the density of the ISM as    
$n_{\rm ISM} = 2$\,cm$^{-3}$. The end of the Sedov-Taylor phase follows
then as $t_{\max}= 13.000$\,yr \citep{TM99}.  For the
magnetic field  we use $B = 1\mu$G and $f_B=1$, if not otherwise
stated.

%%%%%%%%%%%%%%%%%%%%%%%%%%%%%%%%%%%%%%%%%%%%%%%%
\begin{figure}
\includegraphics[width=0.9\linewidth]{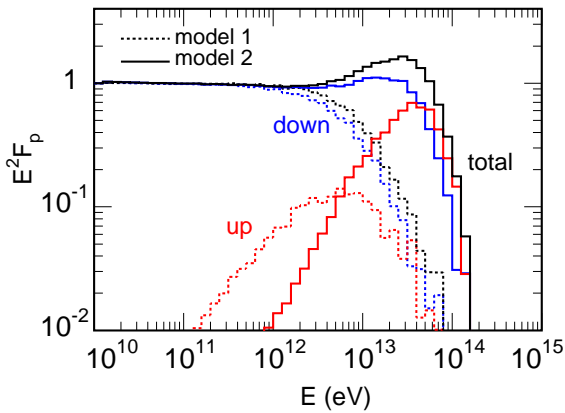}
\caption{
Proton spectra (black) as function of energy for two different injection models
and $f_BB=1\,\mu$G. 
Additionally the contribution of protons staying until $t_{\rm max}$ in the 
up-stream (red) and of protons in down-stream 
region (blue) are shown.
\label{fig:CR}}
\end{figure}
%%%%%%%%%%%%%%%%%%%%%%%%%%%%%%%%%%%%%%%%%%%%%%%%%

%%%%%%%%%%%%%%%%%%%%%%%%%%%%%%%%%%%%%%%%%%%%%%%%
\section{Numerical results}\label{numres}
%%%%%%%%%%%%%%%%%%%%%%%%%%%%%%%%%%%%%%%%%%%%%%%%
In Fig.~\ref{fig:CR}, we show the energy dependence of the proton spectra
in model 1  and 2. Additionally to the total spectra, the contribution
of protons staying at $t_{\rm max}$ in the up-stream region is shown
in red, while the spectra of protons advected to the down-stream are
shown in blue. We do not show the spectra of electrons, 
since they have the same shape as the proton spectra apart from a 
somewhat lower cutoff energy due to synchrotron losses. 
While the total energy spectra  at low energies agree well 
with a $1/E^2$ power-law, changing the injection model leads to large
differences at high energies. The strong dependence of $E_{\max}$ on the 
injection model is expected, since the maximal energy is sensitive to 
how many particles are injected early, when shock acceleration is most 
effective.
However, one may wonder why a bump close to $E_{\max}$ exists in model~2,
while model~1 seems to be well described by an exponential cutoff.

%%%%%%%%%%%%%%%%%%%%%%%%%%%%%%%%%%%%%%%%%%%%%%%%
\begin{figure}
\includegraphics[width=0.9\linewidth]{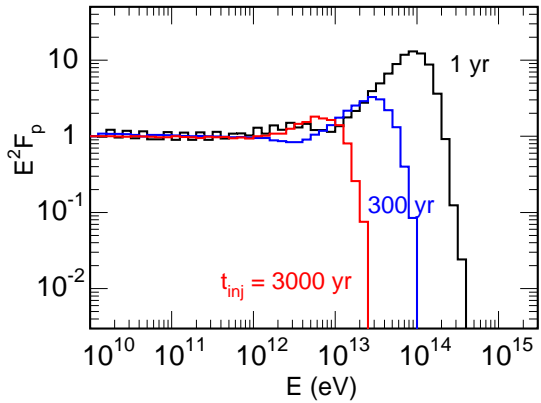}
\caption{
Spectra of cosmic rays  at $t_{\max}$ as function of energy
$E$ for different injection times  $t=1, 300$ and 3000\,yr.
\label{fig:t}}
\end{figure}
%%%%%%%%%%%%%%%%%%%%%%%%%%%%%%%%%%%%%%%%%%%%%%%%%

Apart from energy losses and interactions that do not influence the
proton spectrum for the parameters chosen, deviations from a $1/E^2$ power-law
can be introduced by an additional factor:  The energy spectra of particles 
left behind the shock may be at any time $t$ a scale-invariant
$1/E^2$ power-law up to $E\lsim E_{\max}(t)$, while the total energy of 
particles still participating in acceleration is concentrated around 
$E_{\max}(t)$ as shown by the up-stream component in Fig.~1.
If the latter carry a significant fraction of energy,
as it happens in model 2, a bump in the
total spectrum $E^2 dN/dE$ is visible.
 
To demonstrate this effect, we show in Fig.~\ref{fig:t} the spectra of 
CRs as function of energy $E$, obtained by injecting the same number of  
protons  at different injection times  $t=1$, 300, and 3000\,yr
for model 2.  
Clearly, the peak becomes more pronounced with increasing $E_{\max}$.
The overdensity $\delta N$
of particles contained in the peak is small, but becomes pronounced
in the $E^2 dN/dE$ plot since the associated energy  $\delta E=E\delta N$
is significant.
If this picture is correct, then the CR spectrum of model 1 should also 
develop a peak close to $E_{\max}$ at sufficiently late time $t\gg t_{\max}$.
We confirmed numerically that this is indeed the case.

\begin{figure}
\includegraphics[width=0.9\linewidth]{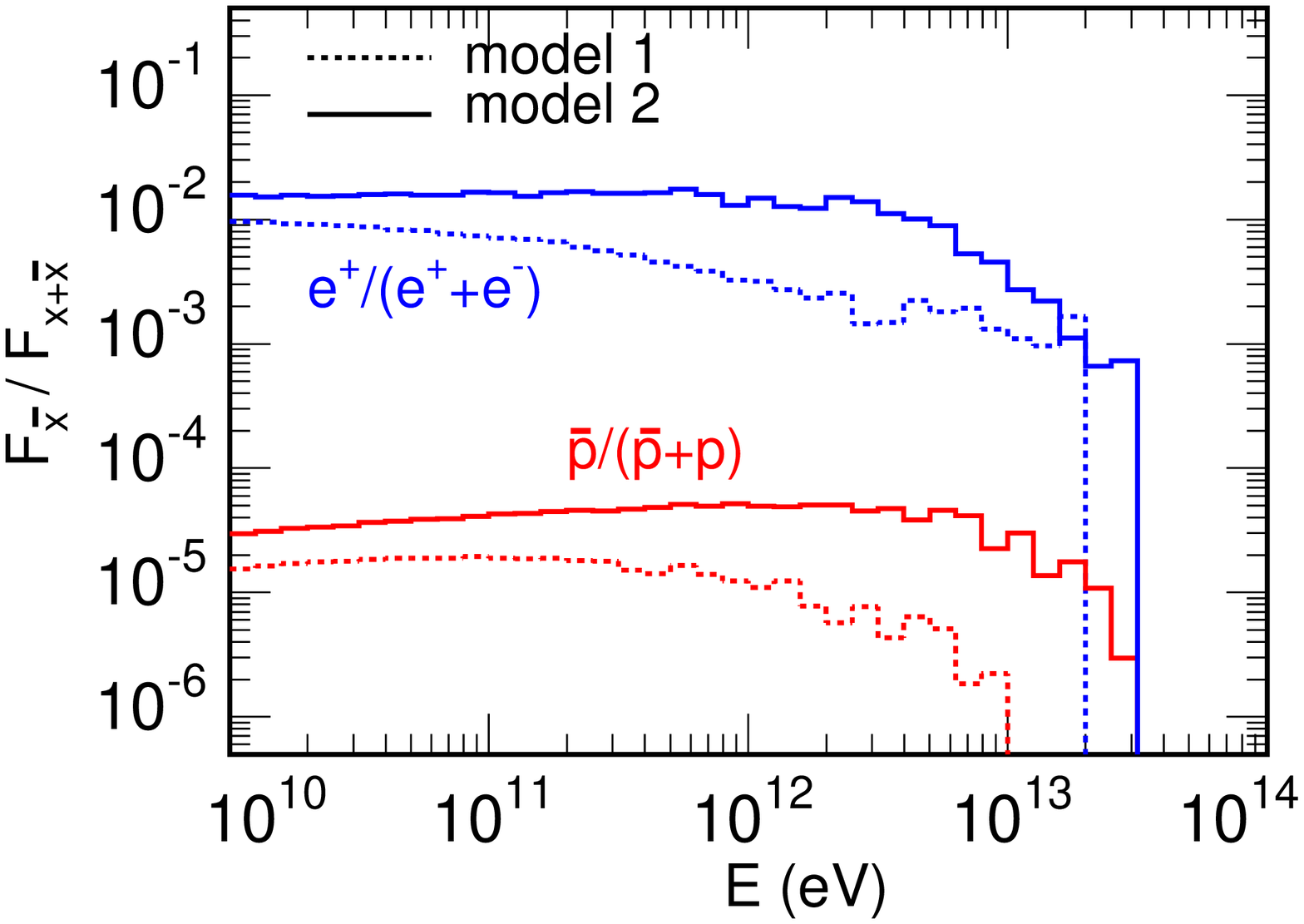}
\caption{
The positron ratio $\re$ (blue) and
antiproton ratio $\rp$ (red) in model 1 (dotted) and 2 (solid).
\label{fig:re}}
\end{figure}

\begin{figure}
\includegraphics[width=0.9\linewidth]{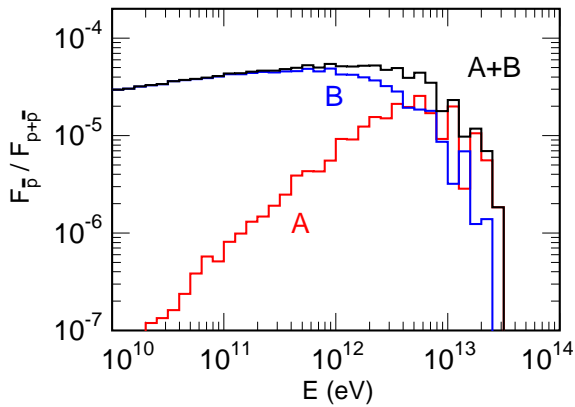}
\caption{
The total flux of antiprotons together with the contribution
A and B in model 2 as function of energy.
\label{fig:rp}}
\end{figure}

%%%%%%%%%%%%%%%%%%%%%%%%%%%%%%%%%%%%%%%%%%%%%%%%%%%%%%%%%%%%%%%%%%%%%%%%%
%{\em Numerical results: Antimatter spectra---}%
%
We switch now to the discussion of the
  secondaries produced. In Fig.~\ref{fig:re} we show their energy
  spectra for the two
injection models considered.
In addition, in Fig.~\ref{fig:rp} we show for injection model~2 and $f_BB=1\mu$G
the antiproton flux split into a part produced in the acceleration
zone (A) and a part produced in the inner part of the SNR (B). More
exactly, we define the contribution~A as all the secondaries that
crossed at least once the shock. This contribution increases fast,
since the time $t_{\rm acc}$ primary protons stay in the acceleration
zone and can interact increases as $t_{\rm acc}\propto D(E)\propto
E$. 
 Moreover,  the inelasticity, i.e.\ the energy
  fraction transferred to {\em all\/} antiprotons is practically
  constant,
   $\langle z_{\bar p}\rangle\ap 0.02$, in the relevant energy range, 
$E_0/\langle z_{\bar p}\rangle\gsim 10^{11}$\,eV. Therefore it does
not influence the shape of the antiproton flux. 
At $E_{\rm b}\ap 2\times 10^{12}$\,eV, the increase of contribution A 
stops, the total flux retains its approximate $E^{-2}$ slope  
and stays small in contrast to the result of \cite{pp}.

%%%%%%%%%%%%%%%%%%%%%%%%%%%%%%%%%%%%%%%%%%%%%%%%
%%%%%%%%%%%%%%%%%%%%%%%%%%%%%%%%%%%%%%%%%%%%%%%%%

%%%%%%%%%%%%%%%%%%%%%%%%%%%%%%%%%%%%%%%%%%%%%%%%
%%%%%%%%%%%%%%%%%%%%%%%%%%%%%%%%%%%%%%%%%%%%%%%%%

How can we understand this behavior and the  maximal value of 
$\rp$? We may assume in a gedankenexperiment that in each 
$pp\to \bar p+X$ interaction the most energetic antiproton 
carries away all the energy, $E_{\bar p}\ap \langle z\rangle E_{p}$ with 
$\langle z\rangle\ap 1$. Then interactions just convert part of the $p$
into a $\bar p$ flux. But since $p$ and $\bar p$ diffuse and are accelerated 
in the same way,  the total $\bar p$ flux is not affected if the $\bar p$
or the parent proton is accelerated. Hence the total flux 
of antiprotons produced in the acceleration zone and inside the SNR,
i.e.\ the sum of A and B, should be simply the proton flux scaled down 
by a constant factor. In particular, the secondary flux of the species $i$ 
is bounded by the proton interaction depth $\tau$ and the 
(spectrally averaged) energy fraction $\langle z_i\rangle$ transferred to $i$.
The maximal conversion rate during the life-time of a SNR is 
with $\sigma^{\rm inel}_{pp}\simeq 30$\,mb as inelastic pp cross section at
100\,GeV given by
$\tau = c\,t_{\max}\,R\,n_{\rm ISM}\,\sigma_{pp}^{\rm inel}\ap 3\times 10^{-3}$. 
The mean energy fraction of antiprotons (plus antineutrons) is 
$\langle z_{\bar p}\rangle\ap 0.02$, so we may expect a maximal 
ratio  of $\rp\sim \langle z_{\bar p}\rangle\,\tau
 \sim 6\times 10^{-5}$. The obtained $\rp$ ratio shown in  
Fig.~\ref{fig:rp} is indeed close to this estimate.

The same discussion applies to the case of positrons, with the sole 
exception that the primary electron flux is scaled down by the factor 
$K_{ep}$ and that the energy fraction transferred to positrons is 
$\langle z_{e^+}\rangle\ap 0.05$. The results of our simulation are shown
in Fig.~\ref{fig:re}, confirming with the maximal value of 
$\re\lsim {\rm few}\;\%$  this simple picture. Note that while
above $\sim 100$\,GeV the ratio $\re$ from SNR starts to be larger than
the conventional prediction using only secondary production on the ISM,
it cannot explain the rise to $\re\ap 10\%$ at 10\,GeV in the PAMELA
data~\cite{PAMELA}.

%%%%%%%%%%%%%%%%%%%%%%%%%%%%%%%%%%%%%%%%%%%%%%%%
\section{Discussion}
%%%%%%%%%%%%%%%%%%%%%%%%%%%%%%%%%%%%%%%%%%%%%%%%

Both the absolute normalization of secondary positron and antiproton spectra
and, more importantly, their spectral shapes obtained disagree 
with the results of \cite{B09,pp,sarkar}. 
In the following, we discuss the reasons for these discrepancies.

%%%%%%%%%%%%%%%%%%%%%%%%%%%%%%%%%%%%%%%%%%%%%%%%%%%%%%%%%%%%%%%%%%%%%%%%%
\subsection{Secondary production}

%%%%%%%%%%%%%%%%%%%%%%%%%%%%%%%%%%%%%%%%%%%%%%%%
\begin{figure}
\includegraphics[width=0.9\linewidth]{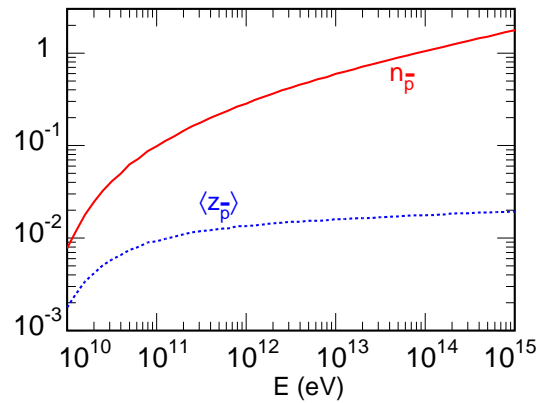}
\caption{
The inelasticity $\langle z_{\bar p}\rangle$ and the multiplicity 
$n_{\bar p}$ of antiproton production as function of the primary
proton energy $E$.
\label{fig:Z}}
\end{figure}
%%%%%%%%%%%%%%%%%%%%%%%%%%%%%%%%%%%%%%%%%%%%%%%%%

One of the main advantages of the Monte Carlo approach is the
  possibility to include interactions and the production of a multi-particle
  final state in an easy and self-consistent way.
  In contrast, the  analytical calculations of \cite{B09,pp,sarkar} 
  involved certain approximations for the  production of
  secondaries. In particular, a constant average energy 
   fraction per {\em single\/} antiproton (or positron) $\xi_i=\langle
  z_i\rangle/\langle n_i\rangle$ was assumed, with
   $\xi_{\bar p}=0.17$ and $\xi_{e^+}=0.05$.

  In reality, it is  the inelasticity,
i.e.\ the energy fraction
\be
\langle z_{\bar p}\rangle\equiv \frac{1}{\sigma_{pp}^{\rm inel}(E)}
\int \!dE'\,\frac{E'}{E}\,\frac{d\sigma ^{p\rightarrow \bar p}(E,E')}{dE'}  
 \ap 0.02
\ee
transferred to {\em all\/} antiprotons, which is practically constant in the
relevant energy range, $E_0/\langle z_{\bar p}\rangle\gsim
10^{11}$\,eV. This behavior is illustrated in Fig.~\ref{fig:Z} where
$\langle z_{\bar p}\rangle$ calculated with our simulation
using QGSJET-II is shown. (A similar result is obtained with SIBYLL.) The same
holds true for the inelasticity of positron production, with $\langle
z_{e^+}\rangle\ap 0.05$.  However,  the
multiplicity of secondaries $n_i$ rises relatively fast
at those energies, as can be seen in Fig.~\ref{fig:Z}.
As a consequence, the average energy fraction per
{\em single\/} antiproton (or positron) $\xi_i$ decreases strongly
with energy. Therefore, assuming a constant  
average energy fraction per {\em single\/} secondary,
$\xi_{\bar p}=0.17$ and $\xi_{e^+}=0.05$, as done in \cite{B09,pp,sarkar},
leads to an overestimate of
$\xi_{\bar p}$ by an order of magnitude for $E_{\bar p}\sim$\,TeV. 
Moreover, the
incorrect use of constant $\xi_i$ results in a wrong energy dependence
of the secondary spectra.

\begin{widetext}

%%%%%%%%%%%%%%%%%%%%%%%%%%%%%%%%%%%%%%%%%%%%%%%%%%%%%%%%%%%%%%%%%%%%%%%%%%
\subsection{Comparison with the stationary picture}

Let us next discuss the differences between our results and the ones 
obtained in Refs.~\cite{B09,pp,sarkar}  that were based on
a stationary picture and a simplified kinetic equation. 
In a stationary approach, it is more convenient to use the shock frame
as reference frame, with the shock
position at $x=0$ and  $u_{1}$ as the flow speed upstream.
The phase space density $f(x,p)$ of CR protons is given by a power-law 
spectrum with exponent $\beta=3R/(R-1)$,
\begin{equation}
 f_{{\rm CR}}(x,p)=K\, p^{-\beta}\,
 \Theta(p_{\max}-p)\,\left\{ \begin{array}{ll}
 e^{xu_{1}/D(p)}, & x<0\\
1, & x>0\end{array} \right. \,.
\label{eq:f_CR}
\end{equation}
Here, we have allowed for a spectral cutoff 
in the primary proton spectrum at $p=p_{\max}$. The energy
distribution of CRs is $N_{{\rm CR}}(E,x)=4\pi p^{2}f_{{\rm CR}}(x,p)$.

We solve the transport equation for secondary cosmic ray  phase space density 
\begin{eqnarray}
u\,\frac{\partial f_{{\rm s}}(x,p)}{\partial x} &=&
\frac{\partial}{\partial x}\left(D(p)\,\frac{\partial f_{{\rm s}}(x,p)}{\partial x}\right)
+\frac{1}{3}\frac{du}{dx}\, p\,\frac{\partial f_{{\rm s}}(x,p)}{\partial p}
+q_{{\rm s}}(x,p)\label{eq:diff-eq}  \\
q_{{\rm s}}(x,p) & = & \frac{1}{4\pi p^{2}}\int\! dE_{0}\; 
N_{{\rm CR}}(E_{0},x)\,\frac{d\sigma^{pp\rightarrow{\rm s}}(E_{0},E)}{dE}\, 
n_{{\rm ISM}}\,c\,,\label{eq:q_s}
\end{eqnarray}
following the same steps as \cite{sarkar} but avoiding any additional
approximations for the source term $q_{{\rm s}}(x,p)$ for secondary particles.

The ratio of the secondary CR flux $J_{{\rm s}}(E)$ produced in the
downstream region to the primary flux $J_{{\rm CR}}(E)$  follows then
using a similar notation as \cite{pp} as
\begin{equation}
 \frac{J_{{\rm s}}(E)}{J_{{\rm CR}}(E)} = 
 n_{{\rm ISM}}\, c\left[A(E)+B(E)\right],\label{ratio}
\end{equation}
where $A(E)$ and $B(E)$ are however given by
\begin{eqnarray}
 B(E)&=&\frac{1}{2}\,R\,t_{{\max}}\int_{E/E_{\max}}^{1}\!
 dz\left.\frac{z^{\beta-3}\,
 d\sigma^{pp\rightarrow{\rm s}}(E_{0},z)}{dz}\right|_{E_{0}=E/z}
 ,\label{eq:B}
\\ A(E) & = & \frac{\beta D(E)}{u_{1}^{2}\:E}\left[
 \int_{0}^{E}\! dE_{0}\:\sigma_{pp}^{{\rm inel}}(E_{0})\,
 I(E_{0},1) + \int_{E}^{E_{\max}}\! dE_{0}\:
 \sigma_{pp}^{{\rm inel}}(E_{0})\,
 I\!\left(E_{0},\frac{E}{E_{0}}\right)\right] \,,
 \label{eq:A}
\end{eqnarray}
with
\begin{equation}
 I(E_{0},\varepsilon)=
 \int_{0}^{\varepsilon}\! dz\;\frac{1}{\sigma_{pp}^{{\rm inel}}(E_{0})}
 \frac{z^{\beta-2}\, d\sigma^{pp\rightarrow{\rm s}}(E_{0},z)}{dz} 
 \left[R^{2}+\frac{1}{z}\right] \,.
 \label{I}
\end{equation}

%%%%%%%%%%%%%%%%%%%%%%%%%%%%%%%%%%%%%%%%%%%%%%%%
\begin{figure}
\includegraphics[width=0.45\linewidth]{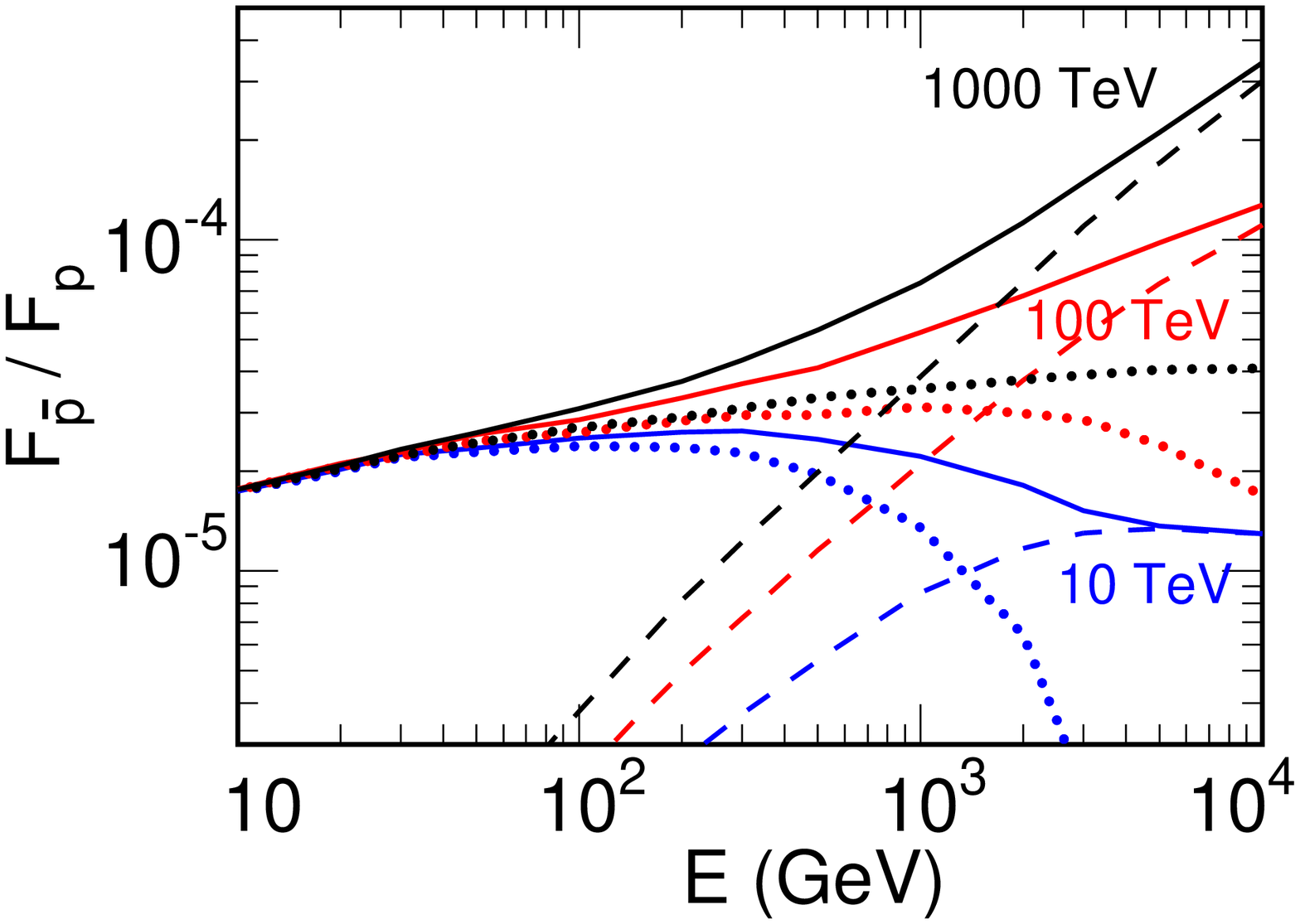}
\includegraphics[width=0.45\linewidth]{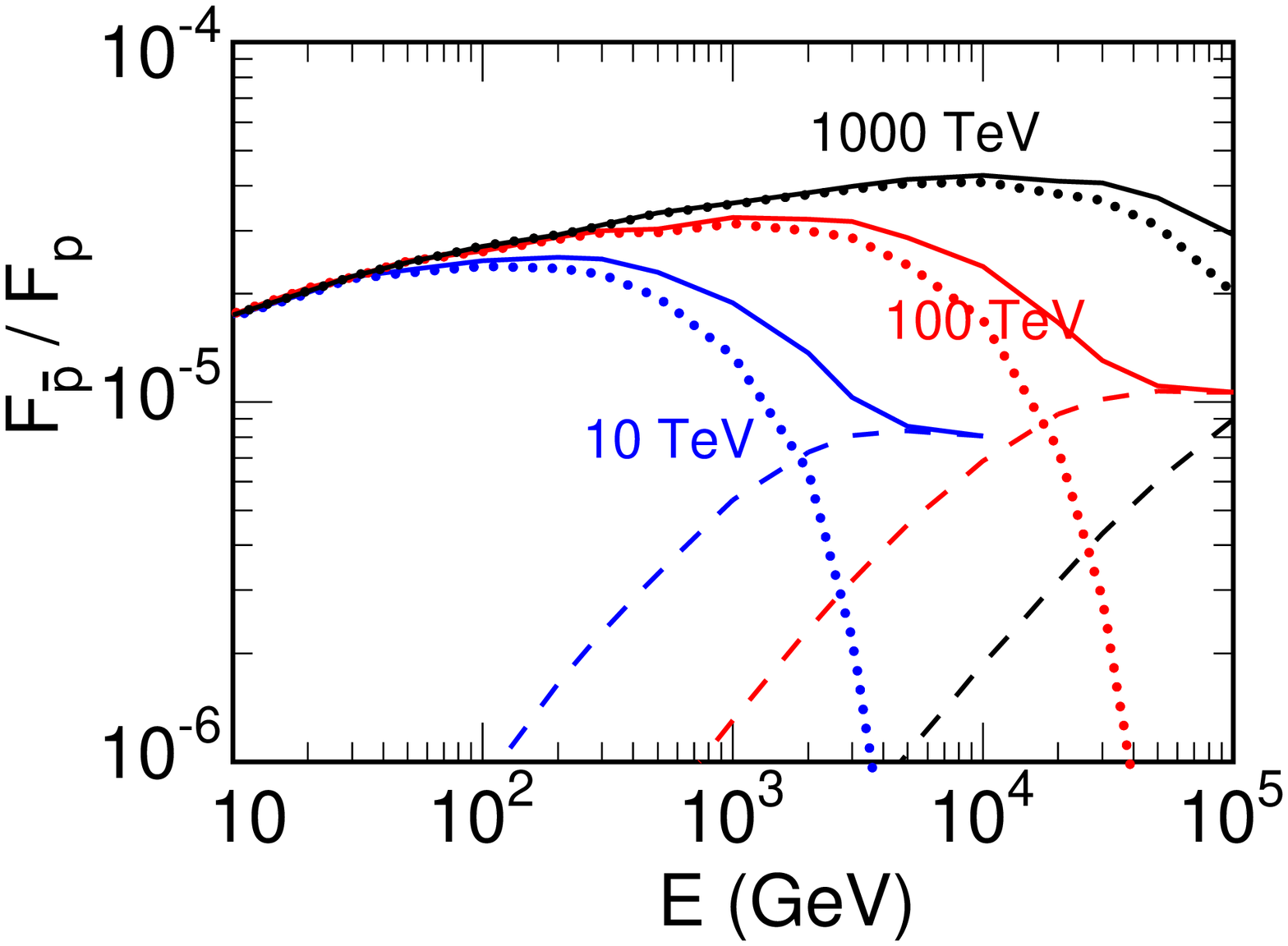}
\caption{The antiproton ratio $\rp$ (solid lines)
for different $E_{\max}$ (as indicated
in the plot) as calculated for a fixed $u_1=10^8$~cm/s (left), and for $u_1$
derived via the relation $D(E_{\max})/u_1^2= t_{\max}/20$ (right)
for the corresponding $E_{\max}$. Partial contributions to the ratio
of the components A and B are shown as dashed and dotted lines, respectively.
\label{fig:stat}}
\end{figure}
%%%%%%%%%%%%%%%%%%%%%%%%%%%%%%%%%%%%%%%%%%%%%%%%%

\end{widetext}

In the left panel of Fig.~\ref{fig:stat}, we show the spectral ratio  
$J_{{\rm s}}(E)/J_{{\rm CR}}(E)$  for the case of secondary antiprotons
as well as the partial contributions $c\,n_{\rm ISM}A(E)$ and $c\,n_{\rm ISM}B(E)$.
We employed in these calculations the same hadronic interaction  
model and the same source parameters ($R=4$, $n_{\rm ISM} = 2$\,cm$^{-3}$, 
$t_{\max}= 13.000$\,yr, $f_B\,B = 1\,\mu$G) as in our MC simulations,
used  a constant speed $u_1=10^8$~cm/s and varied the high energy cutoff 
$E_{\max}$ of the primary CR spectrum. The spectral index of the phase space
density $f$ of primary CRs has been fixed to $\beta=4$.

First, we observe a much stronger dependence of the component A on the choice
of $E_{\max}$ as \cite{pp}. Inspection of
Eqs.~(\ref{eq:A}) and (\ref{I}) shows that the dominant 
contribution to $A(E)$ for $E\ll E_{\max}$ comes from the second integral 
on the r.h.s. of Eq.~(\ref{eq:A}): for small $\varepsilon$ 
$I(E_{0},\varepsilon)\sim \varepsilon$, and the main contribution comes
from the $1/z$ term inside the square brackets of Eq.~(\ref{I}). 
As a result, assuming the
scaling picture\footnote{In reality, Feynman scaling is  broken 
by the logarithmic rise of $\sigma_{pp}^{{\rm inel}}(E_{0})$ and by 
a small power-law like rise with energy of the central 
rapidity density of secondary 
particles.},
$A(E)$ diverges logarithmically, $A(E) \sim D(E) \ln(E_{\max}/E)$,
for $E_{\max}\rightarrow \infty$.

The authors of Refs.~\cite{B09,pp} replaced the factor $1/z$ inside the
integral of Eq.~(\ref{I}) by its ``characteristic value'' $1/\xi$
in front of the integral.
 Such a procedure would be permitted
 if $1/z$ was replaced by its correct (energy-dependent) average with respect 
to the integrand of Eq.~(\ref{I}). 
The same procedure of ``hiding'' this 
logarithmic divergence was copied by~\cite{sarkar}.
The incorrect averaging of the factor 
$1/z$ together with the misconception that the energy transferred to a 
{\em single\/} antiproton or positron per interaction is energy independent
are two major flaws in the analysis of~\cite{B09,pp}.

Using the correct expressions, Eqs.~(\ref{eq:A}-\ref{I}),
the spectral shape of secondary cosmic rays can be almost arbitrarily 
modified in the stationary approach, if  $E_{\max}$  is treated as a free
parameter.
In contrast, using the usual relation for  $E_{\max}$ \cite{lagage}
\[D(E_{\max})/u_1^2\sim t_{\max}/20\,,\]
the relative normalization of the A and B components is fixed.
In such a case, the calculated antiproton to proton spectral ratio,
as plotted in the right panel of Fig.~\ref{fig:stat}, agrees qualitatively 
with our MC results and does not have  the steep energy rise predicted in 
Refs.~\cite{B09,pp,sarkar}.

%%%%%%%%%%%%%%%%%%%%%%%%%%%%%%%%%%%%%%%%%%%%%%%%%%%%%%%%%%%%%%%%%%%%%%%%%%%
\subsection{Parameters determining the acceleration process}
 
In Sect.~\ref{numres}, we have discussed only numerical results for constant 
$f_BB=1\mu$G and one may wonder if a ``better'' choice of parameters can 
increase the antimatter fluxes. In particular, the analytical formula
in the stationary approach of \cite{B09,pp,sarkar} seem to imply 
that the contribution A of antiparticles produced inside the acceleration
zone increases 
for weaker diffusion, i.e.\ larger $D$. However, the term $D/v_1^2$
regulating the importance of A limits also via $t_{\rm acc}\propto
D/v_1^2<t_{\max}$
the maximal proton energy. 
Using a constant value $f_B=1/20$ as in~\cite{B09,pp,sarkar}
thus reduces $E_{\max}$  by the same factor.

The relative size of the partial contributions A and B in Fig.~\ref{fig:rp} 
can be understood
considering the relation between the time $t_{\rm acc}$ spent by protons in 
A, their final energy $E\propto t_{\rm acc}$ and thus the interaction
depth $\tau_A$ in A as function of energy, $\tau_A\propto t_{\rm acc}\propto E$.
In particular, it takes all the  life-time $t_{\max}$ of the  SNR to 
accelerate protons to the highest energies, cf. the up-stream component
in Fig.~1. For the component B,
the optical depth $\tau_B$ of the parent proton is  
$\tau_B\propto  (t_{\max}-t_{\rm acc})$,
which  explains why the two components
A and B sum up to an approximately flat spectrum.
On the other hand,   the relative normalization
of the components A and B in the stationary approach of
\cite{B09,pp,sarkar}
has been imposed by hand: treating
 $E_{\max}$ as an external parameter and   increasing it
relative to its natural value given by $t_{\rm acc}=t_{\max}$
one enlarges the relative contribution of~A,
as illustrated in Fig.~\ref{fig:stat}. Indeed, for $E_{\max}=100$\,TeV
and  the parameters used by \cite{B09,pp}, positron and antiproton
energies
$E\sim E_{\max}$ imply as size of the diffusion zone 
$L_{\rm diff}(E)\sim D(E)/v_2\simeq 2$\,kpc  and as
acceleration time $t_{\rm acc}(E)\sim D(E)/v_2^2\sim 10^7$\,yr.

It is noteworthy that in the treatment of \cite{sarkar}
the limitation due to the finite size of a SNR, 
$L_{\rm diff}(E)<L_{\rm diff}^{\max}\equiv v_2\,\tau_{\max}$,
 has been imposed, which significantly reduced both the relative normalization
 and the steepness of the energy-rise of the component~A compared to the
 original treatment of \cite{B09}. Namely, 
$L_{\rm diff}(E)=L_{\rm diff}^{\max}$ was used for 
$E>E_{\rm break}$, with $E_{\rm break}$ defined by the condition
$L_{\rm diff}(E_{\rm break})=L_{\rm diff}^{\max}$.
However, a more severe constraint on the contribution of the acceleration
zone to secondary antiparticle spectra comes from  the finite life-time of 
a SNR, which was not accounted for fully in the analytic treatments 
of \cite{B09,pp,sarkar}.

One may try to justify the   approach  of  \cite{B09,pp,sarkar} 
as a method to account in an effective way for an amplification
and damping of the magnetic field. Namely, one can assume that the spectra
of primary protons are pre-formed  in the early phase, when
magnetic fields are strongly amplified by
  non-linear effects~\cite{B,B4}, while the production (and re-acceleration) of 
  secondary antiparticles is dominated by the contribution of the 
  Sedov-Taylor phase, when   magnetic fields are damped.

\begin{figure}
\includegraphics[width=0.9\linewidth]{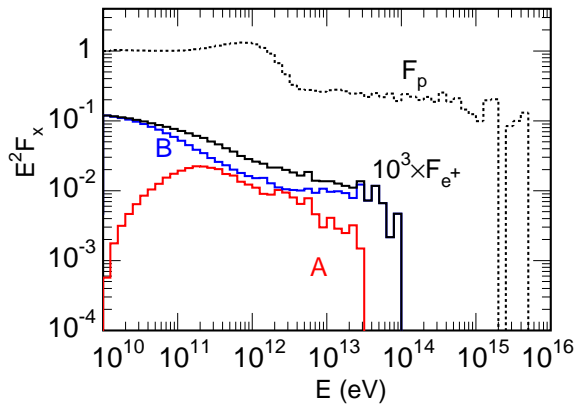}
\caption{
Spectra of cosmic ray protons and positrons (scaled up by a factor 1000)
together with the partial contributions
A and B in model 2  
for a time-dependent diffusion coefficient.
\label{fig:n}}
\end{figure}

We  test this suggestion by considering a
simple toy model for a time-dependent magnetic field: 
assuming it to be strongly amplified in the early phase,
 with $f_BB = 100\:\mu$G before
 the transition to the  Sedov-Taylor phase at $t_\ast=240$\,yr, 
 and using  $f_BB = 1/20\:\mu$G at $t>t_\ast$.
In Fig.~\ref{fig:n}, we show for this case the proton and positron
spectra using the 
injection model 2. Protons that were injected early are accelerated up to 
few$\;\times 10^{15}$\,eV, while the bulk of CRs injected when the turbulent
 magnetic field is damped has a cutoff  around $10^{12}$\,eV.
The contribution A to the positron
flux saturates at $E\sim {\rm few}\;\times 10^{11}$\,eV, i.e.\ at
the energy expected for  $f_BB = 1/20\:\mu$G. In contrast to Fig.~\ref{fig:rp},
the component B dominates now the high-energy end of the positron flux.
It is easy to see the physical picture behind these results:
the damped magnetic field is unable to retain pre-accelerated protons of 
energies $E>10^{12}$\,eV in the vicinity of the shock front, which  thus
escape from the acceleration zone during the transition 
$f_BB/\mu{\rm G}=100\to 1/20$ and escape far upstream  or
are advected downstream.
As a consequence, the re-acceleration of secondary antiparticles of energies
$E>10^{11}$\,eV is only possible in the very beginning of the second phase,
 while at later times the antiparticles produced by CRs escaping up-stream 
and downstream  are not longer accelerated. 
Although different and more realistic scenario for the dumping of the magnetic turbulence
can be considered, it is obvious that the qualitative behavior
of the spectra of secondary cosmic rays will not be significantly modified:
A slower damping would prolong the rise of the component A to higher energies
while reducing its overall normalization ($\propto D/v_1^2$). 
It is worth stressing that the corresponding picture is
essentially a non-stationary one and therefore the complicated
interplay of particle escape and re-acceleration 
can not be described properly  in a  stationary approach.

%%%%%%%%%%%%%%%%%%%%%%%%%%%%%%%%%%%%%%%%%%%%%%%%
%%%%%%%%%%%%%%%%%%%%%%%%%%%%%%%%%%%%%%%%%%%%%%%%%

Finally, we stress that the splitting between contribution A and B 
is artificial and depends as well as the 
total flux on the definition of the escape flux: If the diffusion 
coefficient drops above a certain energy and/or outside a sufficiently small
radius $r_{\rm sh}+\delta r$ to a value close to the one typical for the 
Galaxy, then CRs can escape up-stream instead of being confined down-stream.
Clearly, this effect reduces the contribution B. On the
other hand, the bounds  $\re\lsim {\rm few}\;\%$  and   
$\rp\lsim   6\times 10^{-5}$ will become stronger, since also the time for 
interactions in the acceleration zone will be shortened.
Since our maximal values of $\rp$ and $\re$ depend only on $t_{\max}$, which is
lower in escape-limited models than in age-limited ones, we conclude that the 
contribution of SNR to the observed antimatter in CRs does not
lead to pronounced rise of antimatter fractions and is smaller than estimated in
earlier works.

%%%%%%%%%%%%%%%%%%%%%%%%%%%%%%%%%%%%%%%%%%%%%%%%%%%%%%%%%%%%%%%%%%%%%%%%%
\section{Conclusions} 
We calculated the energy spectra of CRs and their secondaries produced 
in a supernova remnant using a simple random walk picture. In contrast
to a previous prediction that the positron fraction $\re$ can rise up to 
40\%--50\% for $K_{ep}=7\times 10^{-3}$, we found that the ratio levels 
off at a few percent. This
value corresponds to the expectation combining the interaction depth 
$\tau \ap 3\times 10^{-3}$ of a proton during the life-time of a SNR with 
the energy fraction $z\sim 0.05$ transferred to positrons. Similarly,
the antiproton ratio $\rp$ does not rise beyond few$\times 10^{-5}$. 
Our results suggest that antimatter production in SNRs cannot explain 
the rise of the positron fraction observed by PAMELA. 
Since a rising antiproton fraction is neither expected from CR interactions 
with the ISM nor from pulsars, measuring 
${\cal F}_{\bar p}/{\cal F}_{\bar p+p}\gg 10^{-4}$
would be therefore a reliable signature for dark matter.

The reason for the discrepancy with earlier works can be
summarized as follows:  
Our MC results for the case of a constant magnetic field
agree qualitatively with the stationary treatment, as
one can see in the right panel of Fig.~6. In both treatments, one does not
observe any pronounced energy rise of the fraction of 
secondary cosmic rays produced by the sources.
On the other hand, the left panel of Fig.~6 illustrates the main reason
for our differences with Ref.~\cite{pp}:
Treating the maximal energy $E_{\max}$ as a free parameter, one
is able to modify arbitrarily the high energy behavior of
the secondary spectra. Additionally, the 
approximation in the solution of the diffusion equation in
Ref.~\cite{pp} influences the behavior of the spectra at 
lower energies, making them approximately independent of $E_{\max}$ -- 
in contrast
to the exact solution. Other differences between our MC
treatment and the analytic one in Refs.~\cite{B09,pp,sarkar} have a much weaker
impact on the obtained results.

While nuclear fragmentation can be also treated in QGSJET, we have not
included yet the nuclear decay chains  required to predict, e.g., the 
boron-carbon ratio. Such ratios will provide  after their measurement 
by the AMS collaboration a tool to clarify if the scenario proposed 
in Ref.~\cite{B09} or the one discussed here is in better agreement with 
data.

%%%%%%%%%%%%%%%%%%%%%%%%%%%%%%%%%%%%%%%%%%%%%%%%%%%%%%%%%%%%%%%%%%%%%%%%%
\acknowledgments
We thank Markus Ahlers, Pasquale Blasi, Philipp Mertsch and Pasquale 
Serpico for criticism.
S.O.\  acknowledges a Marie Curie IEF fellowship and a fellowship
of Norsk Forskningsradet within the programme Romforskning.

%%%%%%%%%%%%%%%%%%%%%%%%%%%%%%%%%%%%%%%%%%%%%%%%%%%%%%%%%%%%%%%%%%%%%%%%%%


\begin{thebibliography}{00}

\bibitem{galprop}
A.~W.~Strong and I.~V.~Moskalenko,
  %``Models for Galactic cosmic-ray propagation,''
  Adv.\ Space Res.\  {\bf 27}, 717 (2001)
  [arXiv:astro-ph/0101068].
  %%CITATION = ASRSD,27,717;%%
%see also \url{http://galprop.stanford.edu/web_galprop/galprop_home.html}

\bibitem{DM}
For a discussion of DM models and references to original works
see 
L.~Bergstr\"om,
  %``Dark Matter Candidates,''
  New J.\ Phys.\  {\bf 11}, 105006 (2009)
  [arXiv:0903.4849 [hep-ph]].
  %%CITATION = NJOPF,11,105006;%%

\bibitem{PAMELA}
  O.~Adriani {\it et al.}  [PAMELA Collaboration],
  %``An anomalous positron abundance in cosmic rays with energies 1.5.100 GeV,''
  Nature {\bf 458}, 607 (2009)
  [arXiv:0810.4995 [astro-ph]].
  %%CITATION = NATUA,458,607;%%

\bibitem{Adriani:2008zq}
  O.~Adriani {\it et al.}   [PAMELA Collaboration],
  %``A new measurement of the antiproton-to-proton flux ratio up to 100 GeV in
  %the cosmic radiation,''
  Phys.\ Rev.\ Lett.\  {\bf 102}, 051101 (2009)
  [arXiv:0810.4994 [astro-ph]].
  %%CITATION = PRLTA,102,051101;%%

\bibitem{Serpico:2008te}
  P.~D.~Serpico,
  %``On the possible causes of a rise with energy of the cosmic ray positron
  %fraction,''
  Phys.\ Rev.\  D {\bf 79}, 021302 (2009)
  [arXiv:0810.4846 [hep-ph]].
  %%CITATION = PHRVA,D79,021302;%%

\bibitem{pulsar}
 A.~K.~Harding and R.~Ramaty, 
%"The Pulsar Contribution to Galactic Cosmic Ray Positrons,"
Proc. 20th ICRC, Moscow, {\bf 2}, 92 (1987);
A.~Boulares, 
%"The nature of the cosmic-ray electron spectrum, and supernova remnant contributions",
Astrophys.\ J.\ {\bf 342}, 807 (1989);
F.~A.~Aharonian, A.~M.~Atoyan and H.~J.~V\"olk,
  %``High energy electrons and positrons in cosmic rays as an indicator of the
  %existence of a nearby cosmic tevatron,''
  Astron.\ Astrophys.\  {\bf 294}, L41  (1995).
  %%CITATION = AAEJA,294,L41;%%

\bibitem{B09}
P.~Blasi,
  %``The origin of the positron excess in cosmic rays,''
  Phys.\ Rev.\ Lett.\  {\bf 103}, 051104 (2009)
  [arXiv:0903.2794 [astro-ph.HE]].
  %%CITATION = PRLTA,103,051104;%%

\bibitem{pp}
P.~Blasi and P.~D.~Serpico,
  %``High-energy antiprotons from old supernova remnants,''
  Phys.\ Rev.\ Lett.\  {\bf 103}, 081103 (2009)
  [arXiv:0904.0871 [astro-ph.HE]].
  %%CITATION = PRLTA,103,081103;%%


\bibitem{SNR}
S.~P.~Reynolds,
%``Supernova remnants at high energies,''
Ann.\ Rev.\ Astr.\ Astrophys.\ {\bf 46}, 89 (2008).


\bibitem{BC}
P.~Mertsch and S.~Sarkar, 
  %``Testing astrophysical models for the PAMELA positron excess with cosmic ray
  %nuclei,''
  Phys.\ Rev.\ Lett., 103, 081104 (2009)
  [arXiv:0905.3152 [astro-ph.HE]].
  %%CITATION = PRLTA,103,081104;%%


\bibitem{piran}
N.~J.~Shaviv, E.~Nakar and T.~Piran,
  %``Natural explanation for the anomalous positron to electron ratio with
  %supernova remnants as the sole cosmic ray source,''
  Phys.\ Rev.\ Lett.\  {\bf 103}, 111302 (2009)
  [arXiv:0902.0376 [astro-ph.HE]].
  %%CITATION = PRLTA,103,111302;%%


\bibitem{coocon}
R.~Cowsik and B.~Burch,
  %``Positron fraction in cosmic rays and models of cosmic-ray propagation,''
  Phys.\ Rev.\  D {\bf 82}, 023009 (2010).
  %%CITATION = PHRVA,D82,023009;%%


\bibitem{sarkar} 
M.~Ahlers, P.~Mertsch and S.~Sarkar,
  %``On cosmic ray acceleration in supernova remnants and the FERMI/PAMELA
  %data,''
  Phys.\ Rev.\  D {\bf 80}, 123017 (2009)
  [arXiv:0909.4060 [astro-ph.HE]].
  %%CITATION = PHRVA,D80,123017;%%


\bibitem{TM99}
J.~K.~ Truelove and Ch.~F.~McKee, 
%``Evolution of Nonradiative Supernova Remnants,''
Astrophys.\ J.\ Suppl.\ {\bf 120}, 299 (1994).

\bibitem{qgsjet}
  S.~Ostapchenko,
  %``QGSJET-II: Towards reliable description of very high energy hadronic
  %interactions,''
  Nucl.\ Phys.\ Proc.\ Suppl.\  {\bf 151}, 143 (2006);
%  [arXiv:hep-ph/0412332].
  %%CITATION = NUPHZ,151,143;%%
  Phys.\ Rev.\  D {\bf 74}, 014026 (2006).


\bibitem{sibyll1.7}
E.-J.~Ahn {\it et al.}, 
 Phys.\ Rev.\  {\bf D  80}, 094003 (2009).

\bibitem{B}
A.~R.~Bell and S.~G.~Lucek, 
Mon.\ Not.\ R.\ Astron. Soc. {\bf 321}, 433 (2001).
%MNRAS {\bf 321}, 433 (2001).

\bibitem{B4}
A.~R.~Bell, 
Mon.\ Not.\ R.\ Astron. Soc. {\bf 358}, 181 (2004).

\bibitem{in1}
M.~A.~Malkov and H.~J.~V\"olk, 
%"Theory of ion injection at shocks."
Astron.\  Astrophys.\ {\bf 300},  605 (1995).


\bibitem{in2}
  V.~S.~Ptuskin and V.~N.~Zirakashvili,
  %``On the spectrum of high-energy cosmic rays produced by supernova  remnants
  %in the presence of strong cosmic-ray streaming instability and  wave
  %dissipation,''
  Astron.\ Astrophys.\  {\bf 429}, 755 (2005)
  [arXiv:astro-ph/0408025].
  %%CITATION = AAEJA,429,755;%%


\bibitem{lagage}
P.~O.~Lagage and  C.~J.~Cesarsky, 
  Astron.\ Astrophys.\  {\bf 147}, 127 (1985).

\end{thebibliography}
\end{document}